\documentclass[preprint2]{aastex}

\received{}
\accepted{}

\slugcomment{To be submitted to the Astrophysical Journal}  

\shorttitle{The Anomalous Arms of NGC 4258}
\shortauthors{Wilson, Yang \& Cecil}

\begin{document}

\title{Chandra Observations and the Nature of the Anomalous Arms of NGC 4258 
(M 106)} 

\author{A. S. Wilson\altaffilmark{1}, Y. Yang} 
\affil{Astronomy Department, University of Maryland, College Park,
MD 20742; wilson@astro.umd.edu, yyang@astro.umd.edu}

\and

\author{G. Cecil}
\affil{Department of Physics \& Astronomy, University of North Carolina,
CB 3255, Chapel Hill, NC 27599-3255; cecil@physics.unc.edu}


\altaffiltext{1}{Adjunct Astronomer, Space Telescope Science Institute,
3700 San Martin Drive,
Baltimore, MD 21218; awilson@stsci.edu}


\begin{abstract}

This paper presents high resolution X-ray observations with Chandra of NGC 4258
and infers the nature of the so called ``anomalous arms'' in this galaxy. The
anomalous arms dominate the X-ray image; diffuse X-ray emission from the
``plateaux'' regions, seen in radio and H$\alpha$ imaging, is also found.
X-ray spectra have been obtained at various locations along the anomalous
arms and are well described by thermal (mekal) models with kT in the range
0.37 - 0.6 keV. The previously known kpc-scale radio jets are surrounded
by cocoons of hot X-ray emitting gas for the first 350 pc of their length.

The radio jets, seen in previous VLBA and VLA observations, propagate
perpendicular to the compact nuclear gas disk (imaged in water vapor maser
emission). The angle between the jets and the rotation axis of the galactic
disk is 60$^{\circ}$. The jets shock the normal interstellar gas along the
first 350 pc of their length, causing the hot, X-ray emitting cocoons noted
above. At
a height of z = 175 pc from the disk plane, 
the jets exit the normal gas disk and then
propagate though the low density halo until they reach
``hot spots'' (at 870 pc and 1.7 kpc from the nucleus), which are seen in
radio, optical line and X-ray emission. These
jets must drive mass motions into the low density halo gas. This high velocity
halo gas impacts on the dense galactic gas disk and shock heats it along and
around a ``line of damage'', which is the projection of the jets onto the
galactic gas disk as viewed down the galaxy disk rotation axis. However,
because NGC 4258 is highly inclined ($i$ = 64$^{\circ}$), the
``line of damage'' projects on the sky in a different direction to the jets
themselves. We calculate the expected p.a. of the ``line of damage'' on the
sky and find that it coincides with the anomalous arms to within 2$^{\circ}$.
It is, therefore, proposed that the anomalous arms, which are known to lie in
the galactic disk, represent disk gas which has been shocked by mass motions
driven by the out-of-plane radio jets. In the inner ($<$ 1 kpc from the
nucleus) regions of the galactic gas disk, the disk gas is sufficiently
dense and tightly bound that it is heated to X-ray emitting temperatures, but
not blown out of the disk.
Thus, in this inner region, the anomalous arms are straight. Further away
from the nucleus ($>$ 1 kpc), some of the disk gas is blown out of the disk
plane towards the halo on the opposite side of the disk from the relevant
radio jet; this effect causes the arms to curve and is also responsible
for the so called ``plateaux'' emission. An alternative is that the jet 
directions were different in the past, so they projected onto the disk in
different directions. Our picture accounts for: (1) the diffuse
character of the anomalous arms, 2) the inferred ionization of the optical
line-emitting gas by shock waves, (3) the angle between the anomalous arms and
the radio jets, (4) the sharp brightness gradients along the outer edges of the
anomalous arms (these edges represent the standing shock
where the high velocity, jet-driven
halo gas meets the disk), (5) the existence of the ``plateaux'', and (6) the 
wide range of radial velocities observed in the ``plateaux''.

\end{abstract}


\keywords{galaxies: active -- galaxies: individual (NGC 4258 (M106))
-- galaxies: jets -- galaxies: nuclei -- galaxies: Seyfert -- X-rays: galaxies}


%

\newpage
\section{Introduction}

The ``anomalous arms'' of NGC 4258 were discovered through H$\alpha$ imaging
by Court\`es \& Cruvellier (1961). These authors noted their unusual diffuse
and
amorphous
appearance, which contrasts strongly with the knotty structure
(from HII regions ionized by hot stars) of normal spiral arms. Later radio
imaging
by van der Kruit, Oort \& Mathewson (1972) showed that the arms are strong radio
emitters, and these authors suggested that the arms are somehow powered by 
ejection
from the galactic nucleus. Since these early works, numerous 
optical (e.g. Burbidge, Burbidge \& Prendergast 1963; van der Kruit
1974; Ford et al. 1986; Martin et al. 1989; Dettmar \& Koribalski 1990;
Rubin \& Graham 1990; Cecil,
Wilson \& Tully 1992, hereafter CWT; Cecil, Morse \& Veilleux 1995; Cecil et 
al. 
2000, hereafter C2000) and radio (e.g. van Albada \& van der Hulst 1982; Hummel,
Krause \& Lesch 1989;
C2000; Hyman et al. 2001) studies have been made of the anomalous arms.

X-ray emission from the SE anomalous arm was discovered by the Einstein HRI
(CWT).
NGC 4258 was imaged with both the PSPC (Pietsch et al.
1994)
and the HRI (Cecil, Wilson \& de Pree 1995, hereafter CWP) aboard
ROSAT (see also
Vogler \& Pietsch 1999). The anomalous arms dominate the HRI
image, although more
diffusely spread
X-ray emission is also
detected. The PSPC spectra suggest that the X-ray emission of the
anomalous arms has a
thermal origin
with kT $\simeq$ 0.3 keV. X-ray observations with ASCA 
and Beppo-Sax have revealed a hard,
heavily-absorbed, power-law component which coincides with the galactic nucleus,
as
well as the extended, softer thermal emission (Makishima et al. 1994; Reynolds,
Nowak \&
Maloney 2000; Fiore et al. 2001).

There is strong evidence that the anomalous arms lie in the disk of NGC 4258
(e.g. van der Kruit 1974; van Albada \& van der Hulst 1982; Hummel, Krause
\& Lesch 1989; Plante et al. 1991; CWT), as follows. (i) the anomalous
arms extend exactly to the edge of the HI 21 cm distribution; (ii) gaps in
HI 21 cm emission are observed at the positions of the principal anomalous
arms; (iii) the mean velocity of ionized gas along the midline of the SE
anomalous arm tends to track that of the adjacent disk; iv) if the
anomalous arms were out of the plane, one would expect to see a large rotation
measure from the arm on the far side of the disk, due to Faraday rotation by
magnetised gas in the plane of the disk. This is not the case. Instead, the
rotation measure increases in the same way when going inwards along both
anomalous arms; (v) in some parts of the arms, the H$\alpha$ emission seems to
originate from a channel or tunnel between elongated molecular clouds seen
strongly 
in CO emission. These molecular clouds must, of course, lie in the disk; and
vi) the 
densities required for the thermal emission in both optical emission lines
and X-rays are much too high for the halo of the galaxy. We shall thus
assume that the anomalous arms are in the disk plane of NGC 4258.

Many models, all involving ejection from the nucleus in some form or other,
were proposed for the anomalous arms in the 1970s and 1980s (e.g.
van der Kruit, Oort \& Mathewson 1972; van Albada 1978; Icke 1979; Sofue 1980;
Sanders 1982). Models with the arms out of the disk plane (e.g. Sofue 1980;
Sanders 1982) may be excluded in view of their conflict with the 
observational data. More recently, most authors have argued that the arms
represent jets emitted into the galactic disk (e.g. Ford et al. 1986; Martin et
al. 1989; CWT). Discoveries in the last few years confirm that
the nucleus of NGC 4258 is active and emits radio jets. These 
discoveries include: a) a nuclear accretion disk (seen in water vapor maser
emission) of radius
$\simeq$ 0.3 pc
in Keplerian rotation about a putative black hole of mass 3.9 $\times$ 10$^{7}$
M$_{\odot}$
(Miyoshi et al. 1995; Herrnstein et al. 1999); b) elongated radio continuum 
emission on 0.01 - 3 pc scales that
aligns closely with both
the rotation axis of the accretion disk and the jet-like radio emission seen
on the kpc scale in VLA maps (Herrnstein et al. 1997; C2000); and c) Mach disk
and bow
shock structures (seen in radio
continuum and optical emission lines) that
represent
the ends of the currently active jets
and are located 870 pc S and 1.7 kpc N of the nucleus
(C2000). 

However, models in which the arms represent jets or plasmoids ejected {\it into
the disk} have serious problems. First, the radio jets are in a different
direction (on the sky) 
to the inner anomalous arms observed in H$\alpha$ and X-rays
(Section 3.2). Second, models in which the jets are emitted into the galactic 
disk have very large energy requirements since much dense interstellar gas must
be pushed aside (cf. van Albada \& van der Hulst 1982). Third, and most
important, it is very hard
to see how such jets could be confined to a thin galactic gas disk while
propagating to the observed distances of $\simeq$ 8 kpc (4$^{\prime}$) from the 
nucleus.

Previous X-ray observations (described above) have limited spectral and spatial
resolutions,
which
have precluded a detailed study of the X-ray emission from the jets.
The ROSAT HRI images have a resolution of $\simeq$ 5$^{\prime\prime}$ (FWHM) 
but virtually
no spectral resolution, while the ROSAT PSPC data have worse spatial resolution
(25 - 50$^{\prime\prime}$) and limited spectral resolution. The ASCA 
observations have
arc minute spatial resolution and spectral resolution characteristic of CCD
detectors
($\sim$ 100 - 150 eV). The Chandra X-ray Observatory provides sub arc second
spatial
resolution and thus allows the jets to be
investigated at similar resolution to VLA 
radio and ground-based optical studies. 
The 
spectral resolution from direct imaging with the CCDs is similar to that 
attained with
ASCA.

In this paper, we present a first observation of NGC 4258 with Chandra. Although
the
exposure time is short ($\simeq$ 14.0 ksec), 
the data allow a detailed comparison
between
the X-rays and
arc second resolution optical and radio images, and a crude investigation
of the
variation of
the X-ray spectrum along the 
anomalous arms. We also infer the nature of the anomalous arms and
show how they are related to the jets seen at radio wavelengths.
We assume a distance of 7.2 Mpc (Herrnstein
et al.
1999) to NGC 4258, so 1$^{\prime\prime}$ = 35 pc,
and adopt a Galactic hydrogen column towards NGC 4258 of
N$_{\rm H}$ (Gal) = 1.19 $\times$ 10$^{20}$ cm$^{-2}$ (Murphy et al. 1996).

Section 2 describes the Chandra observations and their reduction, while Section
3 presents the results of these observations. In Section 4, we show that the
anomalous arms represent
gas in the galactic disk which has been shocked by the out-of-disk radio jets. 
Section 5 gives concluding remarks.

\section{Chandra X-ray Observations and Reduction}

NGC 4258 was observed with the Chandra X-ray Observatory on April 17 2000
using the Advanced CCD Imaging Spectrometer (ACIS) spectroscopic array, which 
provides excellent
spatial resolution ($\le$ 1$^{\prime\prime}$) with medium
spectroscopic resolution ($\simeq$ 130 eV FWHM at 1 keV).
The default frame time of 3.2 secs was used.
The nucleus was centered
20$^{\prime\prime}$ in the --Y direction from the location of best focus on
chip S3. The data were inspected for background
flares and times of bad aspect, but none was found. A new level 2 events file
was made from the events file supplied by
the Chandra Science Center. Response matrix files and ancillary response files
were made using calibration data taken with the chip at -120C, the temperature
at the time the observations were made. The total useful integration time
after all filtering was 14.0 ksec.
Data were analysed using version 1.1.5
of the ciao software and
version 11.0 of XSPEC.

In the presence of large scale X-ray emission, it is difficult to determine the
level
of the background. We therefore used a compilation of observations of
relatively blank fields, from which discrete sources have been
excised. For a particular source extraction region, a corresponding
background region is taken from this compilation. The two regions have
the same physical location on the S3 chip. The background images and
software$\footnotemark$ of Maxim Markevitch were used for this
procedure. The background is very low with most pixels having zero counts.
\footnotetext{Available at
http://hea-www.harvard.edu/$\sim$maxim/axaf/acisbg/}

\section{Results of X-ray Observations}

\subsection{The nucleus}

The region around the nucleus of NGC 4258 is shown in Fig. 1. Two compact
sources,
separated by 2\farcs5 (87 pc), are apparent. The position (J2000) of the
stronger 
source to the NE is $\alpha$ = 12$^{h}$ 18$^{m}$ 57\fs50,
$\delta$ = +47$^{\circ}$ 18\arcmin\ 14\farcs2. This position coincides with both
the radio continuum nucleus
(as measured from the $\lambda$20cm 1\farcs3 resolution
radio continuum map of C2000: $\alpha_{\rm rc}$ = 12$^{h}$ 18$^{m}$ 57\fs515,
$\delta_{\rm rc}$ = +47$^{\circ}$ 18\arcmin\ 14\farcs36 (J2000)) and the nuclear
H$_{2}$O maser source ($\alpha_{\rm m}$ = 12$^{h}$ 18$^{m}$ 57\fs510 
$\pm$ 0\fs004, $\delta_{\rm m}$ = +47$^{\circ}$ 18\arcmin\ 14\farcs27 $\pm$
0\farcs03
(J2000) - Greenhill et al. 1995).
The nuclear source exhibits a hard
spectrum with a large absorbing
column density. Unfortunately, its spectrum is significantly piled-up
in the present observation,
which prevents us
from obtaining reliable spectral parameters; an observation with a shorter 
frame time is
planned to obtain a reliable nuclear spectrum. 
The spectrum of the weaker source to the SW is well 
described
(reduced $\chi^{2}$ = 0.91 for 5 degrees of freedom) by an
absorbed power law spectrum, with absorbing column
N$_{\rm H}$ = (2.0$^{+1.2}_{-1.1}$) $\times$
10$^{21}$ cm$^{-2}$ and a photon index $\Gamma$ = 1.49$^{+0.50}_{-0.37}$ (the 
errors are 90\% confidence for a single interesting parameter). The observed
0.5 - 4.5 keV flux is
7.0 $\times$ 10$^{-14}$ erg cm$^{-2}$ s$^{-1}$ and the absorption-corrected
luminosity is
5.1 $\times$ 10$^{38}$ ergs s$^{-1}$
if it is associated with NGC
4258, as is very likely given its location. The hard spectrum and X-ray
luminosity suggest that the object is an X-ray binary. Most of the observed
absorption is intrinsic to NGC 4258.

\subsection{The anomalous arms}

Fig. 2 shows a grey scale representation of the Chandra full resolution image of
NGC 4258.
The anomalous arms are seen as
diffuse structures extending to the SE and NW of the nucleus. The diffuse
emission is
better seen in Fig. 3, in which the Chandra image has been smoothed to a 
resolution
of 2\farcs7 FWHM. X-ray emission is seen from the fainter radio-emitting 
regions (called ``plateaux'' by van der Kruit, Oort \& Mathewson 1972)
to the NE
of the SE arm and to the SW of the NW arm. 
Also apparent in the X-ray image is the bifurcation of the
NW arm
$\simeq$ 1$^{\prime}$ from the nucleus, with the main arm continuing to the NW 
and a weaker one
extending almost due W.
 
Fig. 4 shows the relationship between the radio and X-ray structures, both 
images
having a resolution of 2\farcs7 FWHM. The correspondence is particularly good 
in the outer
parts of each arm more than 1$^{\prime}$ (2 kpc) from the nucleus.
The SW and S edges of the SE arm and the NE and N edges of the NW arm are sharp
in both radio and X-ray emission.
An enlargement of the central region is shown in Fig. 5.
In the first 7$^{\prime\prime}$ - 10$^{\prime\prime}$ (250 - 350 pc) of the jet
to the north, the X-rays 
are seen to exhibit a
``two-pronged'' structure, enveloping the radio jet. This morphology suggests
that the
X-rays originate in a hot cocoon around the high velocity material responsible
for the
radio emission. Within 1$^{\prime}$ (2 kpc) 
of the nucleus, the brightest X-ray emission
is aligned
in p.a. $\simeq$ --33$^{\circ}$/147$^{\circ}$, in excellent agreement
with the p.a. $\simeq$ --30$^{\circ}$/150$^{\circ}$ of the H$\alpha$ arms, but
in contrast to the almost N-S
alignment of the
radio emission. Taken together with the close alignment between radio
and X-ray emission further
out, the net effect is that the X-ray arms appear straighter than the radio
arms in lower resolution X-ray images (CWP; Vogler \& Pietsch 1999).
The ``N radio hotspot'' (marked as ``N'' in Fig. 5) 
is located 49\farcs9 (1.7 kpc) N (p.a. --5$^{\circ}$) of the nucleus. 
C2000
identified this feature as the end of the currently active jet in view of
(i) its association
with features suggestive of an oblique bow shock and Mach disk in an HST
H$\alpha$
+ [NII]$\lambda$ 6583 image, and (ii) the detailed kinematics
of high velocity gas in long slit spectra at this location.
Close to the N radio hot spot, there is a compact
X-ray source
with a sharp ``leading'' edge (i.e. to the N) and a slower brightness decline
on the upstream
side (to the S, see Fig. 5). In
the full resolution Chandra image, this source is
elongated by $\simeq$ 3$^{\prime\prime}$ in a NE - SW direction, much like the
the H$\alpha$ + [NII] and radio brightness distributions in the HST and
full resolution VLA images, respectively. This morphology is suggestive of 
thermal X-ray emission from
post bow shock
or entrained jet gas, and supports the idea that these radio, optical and X-ray
features mark the
head of the currently active northern jet. A corresponding, but weaker, feature
(the ``S radio hot spot'', marked ``S'' in Fig. 5) is found in the
radio map 24\farcs5 (860 pc) S (p.a. 178$^{\circ}$)
of the nucleus. The S radio hot spot itself is not detected in X-rays, but
there is a weak
(2 contours in Fig. 5) enhancement of the X-ray emission 3$^{\prime\prime}$
ahead (S) of the 
radio emission, which might represent gas heated by the bow shock. This X-ray
enhancement seems
also to be ahead of the bow-like optical line emission (visible at the
bottom of Fig. 7), consistent with the
latter being post-shock
cooled gas. Further along this same direction, a compact X-ray source is
found 77$^{\prime\prime}$ (2.7 kpc) S (p.a. 176$^{\circ}$) of the nucleus
(marked as ``SX'' in Fig. 4).
The position (J2000.0) of this source is $\alpha$ = 12$^{h}$ 18$^{m}$ 58\fs02,
$\delta$ = +47$^{\circ}$ 16\arcmin\ 57\farcs5. No radio source is detected
at this location. This X-ray source, the nucleus and the N radio hot spot
align to within $\simeq$ 1$^{\circ}$. It is tempting to speculate that this
source
may represent another jet - ISM interaction, but there is no evidence,
other than the alignment, that this X-ray source is associated with NGC 4258.

Fig. 6 compares the large-scale X-ray and H$\alpha$ distributions of NGC 4258.
With the 
exception of a number of
point sources, the normal spiral arms are not seen in X-rays. There is  
an excellent overall correspondence between the X-ray 
and H$\alpha$ emissions of the anomalous arms. In particular, the X-ray and
H$\alpha$ emissions align within 1\arcmin\ of the nucleus, where the radio jet
follows a different (almost N - S) direction, as already noted.

Fig. 7 is a superposition of X-ray contours on an HST H$\alpha$ + [NII]
$\lambda$6583 image 
(from C2000) of the inner regions of NGC 4258. Both images are
dominated by the diffuse emission of the anomalous arms. It is noteable
that the first 5$^{\prime\prime}$ - 6$^{\prime\prime}$ (175 - 210 pc) of the
``two pronged'' X-ray structure to the N of the nucleus (Fig. 5) aligns with
two radial H$\alpha$ + [NII] filaments.
These filaments form the inner part of what
C2000 termed a ``nuclear loop'' (see Fig. 5 of C2000) - a roughly
circular, line-emitting structure open to the north.

We have investigated the variation of the spectrum of the X-ray emission along 
the anomalous
arms. A spectrum was obtained for each of nine regions along the
arms.
All regions show soft X-rays, but the inner regions exhibit, in addition, a hard
X-ray component
above 3 keV. Examination of an image in the 3 - 8 keV band shows that the
emission is
circularly symmetric and centered on the nucleus. We have compared the radial 
distribution of this
emission with that expected from the telescope PSF and found good
agreement. It is thus clear
that this extended hard
emission is not real but results from scattering of nuclear X-rays by the
Chandra mirrors.
We therefore restricted our spectral modelling to the 0.3 - 2.0 keV band to 
avoid contamination
by this hard component.

Each region was modelled by a single, absorbed mekal plasma with the abundance
and absorbing
column allowed to vary.
This model provides an approximate description of the spectra 
($\chi^{2}$/d.o.f.  ranges from 
$<1$ to 1.7). The details of the fits are not believable, but they indicate
temperatures
kT in the range 0.37 to 0.6 keV with no apparent trend with distance from the
nucleus.
At the ends of the arms, the required column density is consistent with
the Galactic column,
but further in
there appears to be absorption intrinsic to NGC 4258.
An example of the spectra is shown in Fig. 8. This represents a
rectangular region centered 64
$^{\prime\prime}$ SE of the nucleus with long dimension (along the arm)
38$^{\prime\prime}$ and short dimension 26$^{\prime\prime}$. Details of the
model are given in the figure caption.
A more detailed discussion of the X-ray spectra of the arms is deferred until a
longer Chandra integration, to be obtained in cycle 2, has been obtained.

\section{The Nature of the Anomalous Arms}
\subsection{The radio jets}

The N and S radio hot spots, located 1.7 kpc and 870 pc, respectively, from the
nucleus align closely with all of: i) the radio jet extending $\simeq$ 700 pc
N and $\simeq$ 400 pc S of the nucleus (cf. Fig. 5); ii) the radio continuum
features seen with VLBI, which extend 0.01 - 0.07 pc from the center of the
maser disk (presumed to be the location of the black hole); and iii) the
projection of the rotation axis of the maser disk on the sky. In particular,
the N and S hot spots lie in p.a. --5$^{\circ}$ and 178$^{\circ}$, respectively,
within 
6$^{\circ}$ of the projected inner disk rotation axis of 
--8$^{\circ}$/172$^{\circ}$ reported by Herrnstein et al. (1999). We conclude
that the maser disk has not changed its axis significantly over the time the
jets have taken to reach the hot spots. Given that the ``N radio hot spot'' is
1.7 kpc from the nucleus, the corresponding jet material left the nucleus at 
least 5.1 $\times$ 10$^{21}$/c secs = 5,700 years ago. If the disk is
precessing, the above noted alignment then implies a precessional period of
$\ge$ 3 $\times$ 10$^{5}$ years. This number was obtained by assuming that
any tilt of the inner regions of the disk over the last 5,700 years is
$<$ 6$^{\circ}$, i.e. this time is 1/60 of a precessional period.

We now define the direction of the jet in three dimensions. Following the
nomenclature of Clarke, Kinney and Pringle (1998; see their Fig. 1), let:
\smallskip
\noindent
$\phi$ be the angle between the jet and our line of sight;
\smallskip
\noindent
$\delta$ be the angle between the projection of the jet onto the plane of
the sky and the apparent major axis of the galaxy disk (also in the plane
of the sky); 
\smallskip
\noindent
$i$ be the inclination of the galaxy disk; and
\smallskip
\noindent
$\beta$ be the angle between the jet and the galaxy disk rotation axis.
\smallskip
\noindent
The angle $\phi$ is presumably equal to the inclination of the maser disk, 
given as 82$^{\circ}$ by Herrnstein et al. (1999). The p.a. of the jet on the
sky is --5$^{\circ}$ and the p.a. of the major axis of the galaxy disk is
146$^{\circ}$ (CWT), so the angle between them $\delta$ = 29$^{\circ}$.
The inclination of the galaxy disk is $i$ = 64$^{\circ}$ (e.g. CWT). The
angle $\beta$ is given by

$$ {\rm cos}~ \beta = {\rm sin}~ \delta~ {\rm sin}~ i~ {\rm sin}~ \phi + {\rm cos}~ i~ {\rm cos}~ \phi \eqno(1)$$

\noindent
so $\beta$ = 60$^{\circ}$, in agreement with C2000. The N jet is tilted
slightly towards us and the S jet slightly away. The NE side is the
far side of the galaxy disk and the SW side the near,
based on both the sense of curvature of the
normal spiral arms plus the observed rotation sense of the galaxy, and the 
pattern of dust obscuration. Thus the N (S) radio jet is on the near (far)
side of the galaxy disk (cf. Fig. 9).

The ``two-pronged'' structure of the X-ray emission enveloping the jet extends
$\simeq$ 250 - 350 pc
from the nucleus to the N (section 3.2, Fig. 5). After this,
the X-ray intensity decreases sharply. A weaker, not obviously ``two-pronged'',
feature extends a similar distance S. As noted above, we believe
these X-rays originate in a hot,
shocked cocoon surrounding the jet. This interpretation is consistent with the
H$\alpha$ + [NII] filaments (Section 3.2, Fig. 7) associated with the two
X-ray prongs to the N of the nucleus. These optical filaments form a 
``nuclear loop'' (C2000), a roughly circular structure open on its N side and
suggestive of gas in the jet-heated cocoon that has cooled to a few $\times$
10$^{4}$K. The H$\alpha$ + [NII] emission is brighter on the western than the
eastern side of the loop (see Fig. 5 of C2000 and the present Fig. 7) 
because the western gas is in the galaxy disk and is,
presumably, denser.

Given the above-defined geometry of the jets, they are at a 
height z = 170 pc above the galactic disk when the X-ray emitting 
cocoons end. We argue
that the X-ray emissions of the cocoons end when the jets exit the normal, 
dense gas disk of NGC 4258. There is only weak X-ray
and H$\alpha$ emission along the rest of
the
jets (as far as the radio hot spots) because the jets extend out of
the plane of the galaxy disk ($\beta$ = 60$^{\circ}$) and the gas density
at these distances (z $\ge$ 170 pc) from the galactic disk is too low for
significant thermal emission.

\subsection{The inner anomalous arms}

It is noteable that the NW extension of the brightest X-rays lies in p.a.
--33$^{\circ}$ and extends NW from the NW prong of the ``two-pronged''
structure discussed above and visible in Figs 5 and 7.
This continuity strongly suggests 
that the X-ray
and H$\alpha$ anomalous arms represent gas in the disk of the galaxy that is,
in some way, connected with the out-of-plane radio jets.

The out-of-plane jets presumably propagate within a low density gas, as 
mentioned above. The jets
must drive mass motions and shock waves into this gas, and these motions will 
propagate in all directions.
The density out of the galaxy plane is too low for strong
X-ray emission, but the high velocity, low density halo gas must collide
with the dense
gas in the galaxy disk and could heat it to X-ray emitting temperatures. This
influence
of the out-of-plane, straight jets on the disk 
will be strongest along the straight
line which represents the locus in the galaxy disk to which the jets are 
closest.
This locus is just the projection of the jets onto the galaxy disk from a
distant viewpoint along the rotation axis of the galaxy disk. However, from
our perspective, the galaxy is highly inclined ($i$ = 64$^{\circ}$) which means
that the projection of this locus on the plane of the sky is tilted w.r.t.
the projection of the jets on the plane of the sky. 

We thus need to derive the angle, $\alpha$, between the projection onto the
plane of the sky of the locus in the galaxy disk to which the
jets are closest and the galaxy disk major axis. This locus is, of course,
a straight line as long as the jet is straight and the galactic
disk is not warped. In terms of the angles
defined earlier:
$$ {\rm cos}~ \alpha = {{{\rm sin}~ \phi~ {\rm cos}~ \delta} \over {({\rm sin}^{2} ~\beta ~{\rm cos}^{2} ~i + {\rm sin}^{2} ~i~{\rm sin}^{2} ~\phi ~{\rm cos}^{2} ~\delta)^{1/2}}} \eqno(2)$$
Substituting the values given earlier gives $\alpha$ = 2\fdg5. Since the p.a.
of the galaxy disk major axis is 146$^{\circ}$, the predicted p.a. of the
projection onto the plane of the sky of the locus in the
galaxy disk to which the jet is closest is 148$^{\circ}$.\footnote{The
projection
onto the plane of the sky of the projection of the jet onto the galaxy disk must
lie between the projection of the jet onto the sky and the galaxy major axis.
Therefore, in the solution of equation (2), $\alpha$ must be positive.}
{\it This p.a. is in excellent agreement with that of the X-ray/H$\alpha$
emission within 2 kpc of the nucleus.} We therefore conclude that the straight
part of the anomalous arms (within 1$^{\prime}$ [2 kpc] of the nucleus)
represents the locus in the galaxy disk to which the jets are closest. As
discussed above, the ``damage'' done by the jets will be greatest along this
line
and the anomalous arms then represent gas in the disk which has been shocked by 
mass motions driven by the out-of-plane jets
(Fig. 9). The optical emission-line ratios indicate that the gas in
the anomalous arms is, indeed, ionized by shocks (e. g. van der Kruit 1974;
Dettmar \& Koribalski 1990; Cecil, Morse \& Veilleux
1995).

The diffuse character of both the X-ray (Fig. 2) and H$\alpha$ (Fig. 7) arms 
supports the above
scenario. A true jet within the galactic disk would be expected to contain
narrow, bright structures at all wavelengths. In contrast, the blast wave
driven into the disk by the out-of-plane jet spreads over a significant width 
in the disk, producing hot gas in a wide, diffuse linear structure without sharp
features.

\subsection{The outer anomalous arms}

The above described situation, in which the arms lie strictly within the disk
plane and represent the projection of the straight out-of-plane jets onto the
disk, must break down at $>$ 1$^{\prime}$ (2 kpc) from the nucleus, because
here the arms are observed to curve in a clockwise sense. As long as the jets
are
straight and the disk is unwarped, the projection
of the jets onto the disk must be
a straight line. Two features of the morphologies of the outer arms seen in 
{\it all wavebands} (i.e. radio continuum, X-rays and optical emission lines)
are crucial at this point:

\noindent
(i) the very sharp brightness gradients seen on the outer edges of the arms
(i.e. the SW and S sides of the SE arm and the NE and N sides of the NW arm).
Images which show these sharp edges well include Figs 4 and 6 of the present
paper, Figs 1 and 2 of van Albada \& van der Hulst (1982) and Fig. 6a of
Hyman et al. (2001). It is noteable that these sharp edges are on the side
of the arms which projects closest to the corresponding out-of-plane radio jet.

\noindent
(ii) the extended, diffuse emissions (``plateaux'') and filaments on the inner
sides of the arms (i.e. the NE and N sides of the SE arm and the SW and S
sides of the NW arm). Images which show these structures well include the
same figures cited for (i), plus the H$\alpha$ images shown as Fig. 19 of CWT
and Fig. 1a and b of C2000. These extended, diffuse emission regions are on the
{\it opposite} side of the arms to the corresponding out-of-plane
radio jet.

Close to the center of the galaxy, the gas in the galaxy disk is dense
and tightly bound gravitationally to the disk plane. Further out, the
gas density decreases and the gas is also less tightly bound. We speculate that
the process of excitation of the outer anomalous arms is the same as 
proposed for the inner arms, namely the impact of mass motions driven down from
the galactic halo by the radio jet. The sharp edges are then naturally accounted
for as shock fronts where the fast moving, light halo gas meets the dense disk
gas; as noted above, the sharp edge is on the side of the anomalous arm
towards the relevant radio jet (Fig. 9).
It may be that the relativistic electrons required for the radio emission
of the outer anomalous arms are accelerated at these shock fronts.
The arms curve because the impact is 
now sufficient to begin to drive gas out of the galaxy disk. The extended
emission (``plateaux'') and filaments represent gas that has been driven
out of the gas disk towards the galaxy halo on the opposite side of the disk to
the relevant radio jet (Fig. 9). This gas is shock ionized, as inferred
from the optical emission-line ratios.

In this hypothesis, we expect a wide range of gas motions in the regions of
diffuse emission; here, the observed Doppler motions should differ from the
normal rotational motion of the galaxy at the projected location on the sky.
These ``anomalous velocities'' should be both positive and negative w.r.t.
normal rotation, because the gas will be driven out of the disk in a range of 
directions w.r.t. the disk normal and the galaxy disk is highly inclined.
Further, some of the gas could be falling back onto the disk, as envisaged
by Rubin \& Graham (1990). Such wide ranges of
velocity have been observed in the diffuse emission (``plateaux'') in
H$\alpha$ by van der Kruit (1974) and Rubin and Graham (1990) and in
HI 21 cm emission by van Albada and Shane (1975). Rubin and Graham (1990) note
that, in their spectrum along p.a. 125$^{\circ}$, the range of velocities in
the SE diffuse emission is larger than that seen in the corresponding
emission to the NW. They ascribe the difference to obscuration of
the diffuse emission on the NW side of the galaxy by the intervening, near
side of the galactic disk, in accord with our model (Fig. 9).

An alternative explanation for the curvature of the outer anomalous arms is that
the jet directions were different in the past, so their projections onto the
disk would also have been different. A progressive change of the jet axes with
time (e.g. through ``disk precession'') would lead to curved jets with a curved
projection onto the stellar disk. In such a picture, the ``plateaux'' would
represent material in the galactic disk which was shocked by the out-of-plane
jets in the past, when the jet directions were different.

\section{Concluding Remarks}
\subsection{X-ray results}

Our Chandra observations have revealed the X-ray structure of the ``anomalous
arms'' of NGC 4258
with the highest resolution to date. The X-rays are found to envelop the inner 
(first 350 pc) of
the northern jet, indicating they originate from a hot, presumably shocked,
cocoon. Further out
(to $\sim$ 2 kpc from the nucleus), the straight 
ridge of the brightest X-rays is rotated
by $\sim$ 30$^{\circ}$ in a clockwise sense w.r.t. the radio jet.
There is a very tight association between
the X-ray and H$\alpha$ emissions of the anomalous arms. In the outermost
regions of detected
X-ray emission, where the arms curve in a trailing sense w.r.t. galactic
rotation, the
X-ray and radio arms also correspond well. The spectrum of the X-rays
from the anomalous arms indicates 
a thermal origin with
temperatures in the range 0.37 to 0.6 keV, with no apparent trend with distance
from the nucleus.

The currently active radio jets project almost N-S on the sky. Both end at well 
defined radio hot spots,
which are associated with structures seen in H$\alpha$ + [NII] images
suggestive
of a Mach
disk and bow shock (C2000). X-rays are detected from the N radio hot spot; the
X-ray morphology reveals a
sharp ``leading'' (i.e. to the N) edge and a slower decline on the ``trailing''
edge, suggestive
of emission from a shock. The S radio hot spot is not detected in X-rays, but
weak enhanced X-ray emission 
is found ahead (S) of the radio peak.

\subsection{The anomalous arms}

The combined radio, H$\alpha$ and X-ray images have allowed us
to elucidate the nature of the anomalous arms. The radio jets in the inner
$\sim$ kpc align perpendicular to the compact, circumnuclear gas disk which
has been imaged in water vapor maser emission. The inclination of this disk is
known, so the 3-dimensional orientation of the jet can be deduced. The angle
between the jet and the rotation axis of the galaxy disk is then found to be
$\beta$ = 60$^{\circ}$.

When the jets initially leave the nucleus, they are surrounded by the dense,
normal, gas disk of the inner regions of this SABSbc galaxy. The jets
shock this gas producing a 
hot cocoon which envelops the jets and emits strongly in X-rays
and optical emission lines, as noted above
(see also Figs 5 and 7).
At a height of z = 175 pc from the galactic disk, the jets
exit the dense galactic disk and then propagate in a low density medium until
they come to the ``N radio hot spot'' and the ``S radio hot spot'', located 1.7 
kpc and 870 pc from the nucleus, respectively (Fig. 5). The existence of a 
compact
X-ray source 77$^{\prime\prime}$ (2.7 kpc) S of the nucleus (Fig. 4), aligned to
within 1$^{^\circ}$ with the jet further in, hints that the jets may
propagate beyond the hot spots.
During this phase, the jets heat, and drive mass motions into,
the surrounding gas but little X-ray emission results because of the low
density of the gas in the galactic halo.

The jets, propagating supersonically at 30$^{\circ}$ to the plane of the
galaxy, drive mass motions and shocks in all directions. The mass motions
propagating
towards the 
plane of the
galaxy meet the dense gas disk and heat it to X-ray emitting
temperatures (Section 3.2).
The greatest ``damage'' done by the jet will be along its projection onto the
galaxy gas disk, as viewed down the disk rotation axis. Since NGC 4258 is
substantially inclined ($i$ = 64$^{\circ}$), however, this
``line of damage'' projects on the sky in a different direction to the jet 
itself. Because the 3-dimensional geometry is fully known, we can calculate the
angle between the jet and the ``line of damage'' projected onto the sky. This
simple calculation shows that the anomalous arms (observed in X-rays and
H$\alpha$) coincide with the expected ``line of damage'' to within 2$^{\circ}$.

The picture that emerges is one of a bi-polar jet inclined at 30$^{\circ}$ to
the galactic gas disk and driving shocks into it. This scenario accounts
naturally for the diffuse structure of the arms since the shocks from the jet
heat up a wide area of the disk. 
Preexisting molecular clouds may be too dense to be heated by these
shocks, so the inner anomalous arms represent hot, lower density regions between
these clouds, as observed (Martin et al. 1989: CWT). The optical line emission
represents gas in the galactic disk which has been shock ionized and has then
cooled (van der Kruit 1974;
Dettmar \& Koribalski 1990; Cecil, Morse \& Veilleux 1995). For this reason,
the X-ray and
H$\alpha$ distributions agree closely.

In the inner regions ($<$ 1 kpc from the nucleus) of the galaxy, the gas
density is high and it is tightly bound gravitationally. The blast waves
driven into the gas heat it to $\le$ 1 keV, but are unable to drive the hot
gas out of the disk. In this region, the ``line of damage'' is a straight line
in the disk, so the anomalous arms appear straight on the sky. Further out
($>$ 1 kpc from the nucleus), the gas in the disk is of lower density and
less tightly bound. The impact of the low density halo gas driven by the jet
onto the disk then suffices to drive the disk gas out of the disk plane.
Each arm then curves away from the radio jet that powers it. An alternative
scenario for the curvature of the anomalous arms would involve changes in the
jet directions over time.

In summary, this scenario accounts for:

\noindent
(i) the diffuse character of the anomalous arms;

\noindent
(ii) the observation that the optical line ratios indicate ionization by
shocks;

\noindent
(iii) the angle between the radio jet and the anomalous arms;

\noindent
(iv) the sharp brightness gradients along the outer edges of the arms;
these edges are on the side of the arm nearest to the relevant radio jet.
The sharp edges represent standing shocks at the interface between the low
density halo gas being driven by the jet onto the gas disk and the gas disk
itself;

\noindent
(v) the diffuse gas and filaments (``plateaux'') on the opposite side of the
disk to each jet. These structures, seen in optical emission lines, X-rays
and radio continuum, are gas which has been pushed out of the disk towards
the halo by the impact of the low density high velocity halo gas driven by the 
jet on the other side of the disk;

\noindent
(vi) the observed wide range of velocities of the diffuse gas and filaments.

A prediction of this model is the existence of extremely hot, high velocity
gas in the two
quadrants of the galaxy halo occupied by the radio jets. X-ray absorption
spectra of quasars in these quadrants could be used to search for this gas.
The unique character of the anomalous arms in NGC 4258 is probably related to
the presence of {\it both} powerful jets (generally seen only in early-type
galaxies) and a dense, galactic gas disk (generally seen only in late-type
galaxies). Our scenario could be evaluated quantitatively by a hydrodynamic
simulation, which would allow the required parameters (e.g. jet power,
halo gas density, disk gas density etc.) to be derived.

This research was supported by NASA through grant NAG 81027 and
by the Graduate School of the University of Maryland through a research
fellowship to ASW. We are grateful to A. J. Young for advice on the Chandra
analysis and to Stephen M. White for a helpful discussion. We also wish to 
thank the staff
of the Chandra Science Center, especially Dan Harris and Shanil
Virani, for their help, and the anonymous referee for useful comments.

\vfil\eject

\clearpage

\figcaption[fig1.ps]
{A contour map of the nucleus of NGC 4258 and its
companion in the 0.3 - 10 keV band at the full Chandra resolution. Contours are
plotted at 2.5, 5, 7.5, 10, 15, 20, 30, 50, 100, 200 and 300 cts pixel$^{-1}$.
\label{Figure 1}}

\figcaption[fig2.ps]
{A grey scale representation of NGC 4258 in the 0.3 -8 
keV band at the full Chandra resolution ($\lesssim$ 1$^{\prime\prime}$).
\label{Figure 2}}

\figcaption[fig3.ps] 
{A grey scale representation of NGC 4258 in the 0.3 - 8
keV band with a resolution of 2\farcs7 (FWHM).
\label{Figure 3}}

\figcaption[fig4.ps] {Superposition of radio (grey scale) and X-ray
(contours) images of NGC 4258. The radio image is at 1.49 GHz and was obtained
with the VLA by Cecil et al. (2000). Both images have a resolution of
2\farcs7 (FWHM). Contours of the Chandra image are plotted at 3, 6, 9, 12,
15, 20, 50, 100, 500, 1000 and 1500 cts pixel$^{-1}$. The compact X-ray source
77$^{\prime\prime}$ S of the nucleus, which aligns with the nucleus and
N radio hot spot, is marked by ``SX'' (Section 3.2).
\label{Figure 4}}

\figcaption[fig5.ps] {Superposition of radio (grey scale) and X-ray
(contours) images in the inner regions of NGC 4258. The radio map and the
X-ray contours are as in Fig. 4. The N and S radio hot spots are indicated
by ``N'' and ``S'', respectively.
\label{Figure 5}}

\figcaption[fig6.ps] {A comparison of X-ray (left) and H$\alpha$ (right, from
C2000) images of NGC 4258. The top left panel is the full resolution Chandra
image, while the lower left is the image convolved to 2\farcs7 FWHM. The two
H$\alpha$ images (top and bottom right) are identical but displayed with
different contrasts. The FWHM resolution of the H$\alpha$ images is
$\simeq$ 1\arcsec (C2000).
\label{Figure 6}}

\figcaption[fig7.ps] {Superposition of the X-ray image with resolution
FWHM = 2\farcs0 (contours) on an HST image in the light of
H$\alpha$ + [NII] $\lambda$6583 (from C2000). Contours are plotted at
3, 6, 9, 12, 120 and 1,200 cts pixel$^{-1}$. North is up and east to the
left. The ticks on the border are separated by 10$^{\prime\prime}$. The
correspondence between the ``two pronged'' X-ray emission extending
5$^{\prime\prime}$ - 6$^{\prime\prime}$ N of the nucleus and the 
H$\alpha$ + [NII] emission of the ``nuclear loop'' is noteable (see text).
\label{Figure 7}}

\figcaption[fig8.ps] {Spectrum of a region, described in the text, along the 
anomalous arm to the SE of the nucleus. The model, shown as the 
histogram, is
a mekal with the column density, N$_{\rm H}$, temperature, T, abundance
and normalisation, K, as free parameters. The model has
N$_{H}$ = 2.8$^{+8.2}_{-2.8}$ $\times$ 10$^{20}$ cm$^{-2}$ (and is thus
consistent with the Galactic column), kT = 0.48$^{+0.05}_{-0.12}$, abundance
0.12$^{+0.05}_{-0.06}$ and K = 3.09$^{+5.22}_{-1.07}$ $\times$ 10$^{-4}$. 
Errors are 90\% confidence for a single interesting parameter. 
K is defined as $10^{-14} \int n_e n_H dV / (4 \pi ((1+z)D_A)^2)$, where
$n_e$ is the electron density (cm$^{-3}$), $n_H$ is the hydrogen density 
(cm$^{-3}$), $V$ is the volume (cm$^{3}$), $z$ the redshift and $D_A$
is the angular size distance to the source (cm).
\label{Figure 7}}

\figcaption[fig9.ps] {Schematic diagram (not to scale) showing the nuclear
maser disk, the radio jets, the radio hot spots, the galactic disk and the
anomalous arms. The N (S) radio jets are on the near (far) side of the
galactic disk. We show in Section 4 that the inner anomalous arms represent the
projection of the jets onto the plane of the galactic disk when viewed down
the rotation axis of the galactic disk. We argue that the anomalous arms
represent dense disk
gas which has been shocked by gas driven down onto the disk by the out-of-plane
radio jets. The arms begin to curve some 1$^{\prime}$ (2 kpc) from the nucleus
because at these larger radii the gas is of lower density and less tightly bound
to the galaxy. The mass motions driven by the jet can then push gas out
of the disk in the opposite direction, causing the extended emission features
(``plateaux'') seen in radio, X-ray and optical line emission. On the side of
the anomalous arms closest to the jets, there is a shock front between the
high velocity, low density halo gas and the dense disk gas; this shock front
is seen as the sharp edges of the anomalous arms on the sides which project 
towards the jets. See Section 4 for further details.
\label{Figure 8}}

\clearpage
\end{document}